\documentclass[preprint,preprintnumbers,showpacs,aps,amssymb]{revtex4}

\usepackage{graphicx}
\usepackage{bm}
\usepackage{amsmath}

\def\calL{{\cal L}}
\def\calM{{\cal M}}
\def\calO{{\cal O}}
\def\calU{{\cal U}}

\def\BZ{{\cal BZ}}
\def\dU{{d_\calU}}

\def\ubar{{\bar u}}

\def\nn{\nonumber}

\begin{document}
\title{Deconstructing unparticles in higher dimensions}
\author{Jong-Phil Lee}
\email{jplee@kias.re.kr}
\affiliation{Korea Institute for Advanced Study, Seoul 130-722, Korea}
\preprint{KIAS-P09002}

\begin{abstract}
Unparticles are realized by deconstruction in higher extra dimensions.
It is shown that in this framework when the scale invariance is broken,
the corresponding spectral function of the unparticle is shifted by an
amount of the breaking scale.
The result strongly supports the conventional ansatz for the spectral function
of the unparticle in the literature.
\end{abstract}

\maketitle
Recently Georgi has proposed a possibility that there is a scale invariant
hidden sector which couples to the standard model (SM) sector very weakly 
\cite{Georgi}.
In a low energy viewpoint, the effect of the hidden sector can be easily
described by an effective theory.
Formally there is an operator $\calO_\calU$ of the hidden sector coupled to
the SM operator $\calO_{SM}$.
Due to the scale invariance of the hidden sector, the scaling dimension of
$\calO_\calU$ is nontrivial.
This unusual behavior is reflected on the phase space of $\calO_\calU$;
it looks like a phase space for a fractional number of particles, 
hence dubbed as "unparticles".
After Georgi, there has been a lot of works on unparticles both in theory
and experiment \cite{Cheung}.
\par
The Banks-Zaks ($\BZ$) theory \cite{BZ} provides a good starting point of 
studying unparticle physics.
They examined the infrared(IR)-stable fixed point of Yang-Mills theories with
massless fermions,
and found that for a proper number of fermions in a certain representation
the theory is chirally invariant and has no mass gap.
At a high scale of $M_\calU$ the $\BZ$ sector interacts with the SM fields.
Below $M_\calU$ the interaction can be described as
\begin{equation}
\frac{1}{M_\calU^{d_{SM}+d_{\BZ}-4}}\calO_{SM}\calO_{\BZ}~
\end{equation}
where $\calO_{\BZ}$ is the $\BZ$ operator with scaling dimension $d_\BZ$
while $d_{SM}$ is the scaling dimension of $\calO_{SM}$.
When the scale goes down by the renormalization flow a scale $\Lambda_\calU$
appears through the dimensional transmutation 
where the scale invariance emerges.
Below $\Lambda_\calU$ the theory is matched with the new operator $\calO_\calU$
onto the above interaction as
\begin{equation}
C_\calU\frac{\Lambda_\calU^{d_{\BZ}-\dU}}{M_\calU^{d_{SM}+d_{\BZ}-4}}
\calO_{SM}\calO_{\calU}~
\end{equation}
where $\dU$ is the scaling dimension of $\calO_\calU$ and $C_\calU$ is the
matching coefficient.
\par
The spectral function of the unparticle is given by the two-point function
of $\calO_\calU$:
\begin{eqnarray}
\rho_\calU(P^2)&=&\int d^4x~e^{iP\cdot x}
\langle 0|\calO_\calU(x)\calO_\calU^\dagger(0)|0\rangle\nn\\
&=&
A_\dU\theta(P^0)\theta(P^2)(P^2)^{\dU-2}~,
\end{eqnarray}
where 
\begin{equation}
A_\dU=\frac{16\pi^2\sqrt{\pi}}{(2\pi)^{2\dU}}
\frac{\Gamma(\dU+\frac{1}{2})}{\Gamma(\dU-1)\Gamma(2\dU)}~,
\end{equation}
is the normalization factor.
The corresponding phase space is 
\begin{equation}
d\Phi_\calU(P)=\rho_\calU(P^2)\frac{d^4P}{(2\pi)^4}
=A_\dU\theta(P^0)\theta(P^2)(P^2)^{\dU-2}\frac{d^4P}{(2\pi)^4}~,
\end{equation}
which in $\dU\to 1$ limit reduces to that of a single massless particle.
\par
But there are still many things to be unveiled about the unparticle.
Among them is the scale invariance breaking.
When unparticles couple to the SM sector, Higgs coupling is very important
because it is relevant.
We confine ourselves to the scalar unparticles for simplicity throughout
the paper.
If the Higgs develops vacuum expectation values (VEV) then a scale is 
introduced in the theory, breaking the scale invariance.
Up to now there is no formal way of describing the unparticles with broken
scale-invariance.
A usual method is to adapt a simple ansatz for the spectral function of 
unparticles.
It says that the spectral function of unparticles $\rho_\calU(P^2)$ is
shifted by the scale-invariance breaking scale $\mu$ as \cite{Fox}
\begin{equation}
\rho_\calU(P^2)\rightarrow\rho_\calU(P^2-\mu^2)~.
\label{ansatz}
\end{equation}
The ansatz shows good "correspondences"; 
in the limit of $\mu\to 0$, $\rho_\calU$ reduces to the usual unparticle 
spectral function, and in the $d_\calU\to 1$ limit, the corresponding
propagator is a free particle propagator of mass $\mu$.
\par
In this paper, it will be shown that (\ref{ansatz}) can be justified in the
framework of deconstruction.
Deconstruction \cite{Stephanov} is a useful method to deal with the unparticles.
In this scheme the unparticle is described by an infinite tower of particles
with vanishing masses.
A continuous spectrum of unparticles is simulated by a descrete sum over
deconstructing states, which comes to an integral in the vanishing mass limit.
One way of explicit realization of deconstruction is to use AdS/CFT 
correspondence to build a 5-dimensional field theory \cite{Terning}.
In this work we try flat $4+\delta$ dimensional theory for deconstruction.
The higher dimensional Kaluza-Klein (KK) states are identified as the 
deconstructing fields.
The main point is that the mass $m_0$ of the higher dimensional field really 
shifts the spectral function of the deconstructing fields.
In the absence of $m_0$, the mass spectrum of $n$-th KK states is $\sim n/R$
where $R$ is the size of extra dimension.
When $R\to\infty$ the infinite tower of KK states do the work of deconstruction.
If $m_0$ appears for some reason (probably via Higgs interaction) in the theory,
then the KK spectrum becomes $\sim\sqrt{m_0^2+n^2/R^2}$,
which can be interperted as the shift of the spectral function by $m_0^2$.
\par
Another good point of this picture is that the scale $m_0$ regulates the
infrared (IR) divergence of the VEV of $\calO_\calU$.
It was already shown that when Higgs couples to unparticles, there needs an
IR cutoff for the VEV of $\calO_\calU$ \cite{DEQ}.
In \cite{DEQ} a scale invariant quartic coupling of Higgs and deconstructing 
fields is introduced for regulation.
When Higgs gets VEV, the scale invariance is broken and the deconstructing
field has "mass gap". 
We identify this mass gap as nothing but $m_0$ of higher dimensional theory.
\par
The unparticle will be realized in this work by deconstruction.
For this purpose it is assumed that there is a hidden sector of scale invariance 
with extra dimensions.
A scalar field in this sector has its KK excitations.
If the size of extra dimension is large enough then the masses of the KK
states is vanishing.
In this limit the infinite tower of this massless KK states is identified as
the deconstructed form of scalar unparticles.
\par
Consider the higher dimensional description of a scalar field.
We work in the flat $D=4+\delta$ dimensions where $x^\mu~(\mu=0,1,2,3)$ is the
4-dimensional coordinate while $y^a~(a=1,\cdots,\delta)$ is the extra
dimensional one. 
It is assumed that the extra higher dimension is compactified on an
$[(S^1\times S^1)/Z_2]^k$ orbifold for $\delta=2k$ and
$[(S^1\times S^1)/Z_2]^k\times (S^1/Z_2)$ orbifold for $\delta=2k+1$ 
\cite{ACD}.
\par
For a higher dimensional scalar field $\phi(x^\mu,y^a)$, 4-dimensional effective
Lagrangian is given by
\begin{equation}
\calL_0(x^\mu)=\int d^\delta y\left(
 \frac{1}{2}\partial_M\phi\partial^M\phi\right)
=
\int d^\delta y\left(\frac{1}{2}\partial_\mu\phi\partial^\mu\phi
-\frac{1}{2}\partial_a\phi\partial^a\phi\right)~,
\label{L0}
\end{equation}
where $M=0,1,\cdots,\delta$.
The Fourier expansion for the extra dimensional component of $\phi$ yields
\begin{equation}
\phi(x^\mu,y^a)=\frac{1}{(2\pi R)^{\delta/2}}\left\{
\phi_0(x^\mu)+\sqrt{2}\sum_{j_1,\cdots,j_\delta}^\infty
\phi_{j_1,\cdots,j_\delta}(x^\mu)
\cos\left(\frac{j_1y_1+\cdots+j_\delta y_\delta}{R}\right)\right\}~.
\end{equation}
Here $R$ is the size of the extra dimension, and we impose the symmetry of
$\phi(x^\mu,y^a)=\phi(x^\mu,-y^a)$.
If we put the Fourier decomposition of $\phi(x^\mu,y^a)$ into Eq.\ (\ref{L0}),
we have
\begin{equation}
\calL_0(x^\mu)=\frac{1}{2}\Bigg\{
\partial_\mu\phi_0\partial^\mu\phi_0
+\sum_{[j]}\partial_\mu\phi_{[j]}\partial^\mu\phi_{[j]}
-\sum_{[j]}\left(\frac{j_1^2+\cdots+j_\delta^2}{R^2}\right)
\phi_{[j]}^2\Bigg\}~,
\end{equation}
where $[j]$ denotes the collective $j_is$, $\{j_1,\cdots,j_\delta\}$.
For a given $j_1^2+\cdots+j_\delta^2\equiv k^2_\delta$, there is a degeneracy 
$D_n$ where $k_\delta/R$ is the $n$-th Kaluza-Klein (KK) mass spectrum.
Let the corresponding $n$-th level field be $\phi_n$.
Then there are $D_n$ number of $\phi_n$s at $n$-th level with mass $k_\delta/R$.
In the limit $R\to\infty$ the mass gap between the KK levels goes to zero.
Fields of infinite tower with vanishing mass gap can be assembled to form an
unparticle operator in a deconstructed form \cite{Stephanov}.
\par
A scalar unparticle operator can be defined by
\begin{equation}
\calO_\calU\equiv\sum_{n=0}^\infty F_n\phi_n~.
\end{equation}
The spectral function of $\calO_\calU$ is then
\begin{eqnarray}
\rho_\calU^0(P^2)&=&
\int d^4x e^{iP\cdot x}\langle 0|\calO_\calU(x)\calO^\dagger_\calU(0)|0\rangle
\nn\\
&=&2\pi\sum_n\delta(P^2-m_n^2)F_n^2~,
\end{eqnarray}
where the superscrip means that the function is defined by $\calL_0$.
The two point function is related to $\rho_\calU$ as
\begin{eqnarray}
D_\calU^0(P)&\equiv&\int d^4x e^{iP\cdot x}
\langle 0|T\calO_\calU(x)\calO^\dagger_\calU(0)|0\rangle \nn\\
&=&
\int\frac{dM^2}{2\pi}\frac{i\rho_\calU^0(M^2)}{P^2-M^2+i\epsilon}\nn\\
&=&
\sum_n\frac{iF_n^2}{P^2-m_n^2+i\epsilon}~,
\end{eqnarray}
where 
\begin{equation}
m_n^2=\frac{k_\delta^2}{R^2}~.
\end{equation}
In the limit of $R\to\infty$ the spectral function should be matched onto
that of the unparticle
\begin{equation}
\rho_\calU^0(P^2)\rightarrow A_{d_\calU}\theta(P^0)\theta(P^2)(P^2)^{d_\calU-2}~.
\label{rho0match}
\end{equation}
\par
Now suppose that the $\phi$ field gets mass term in (\ref{L0}).
Obviously this term breaks the scale invariance.
The effective Lagrangian becomes
\begin{eqnarray}
\calL(x^\mu)&=&\int d^\delta y\left(
 \frac{1}{2}\partial_M\phi\partial^M\phi-\frac{1}{2}m_0^2\phi^2\right)\nn\\
&=&
\frac{1}{2}\Bigg\{
\partial_\mu\phi_0\partial^\mu\phi_0-m_0^2\phi_0^2\nn\\
&&
+\sum_{[j]}\partial_\mu\phi_{[j]}\partial^\mu\phi_{[j]}
-\sum_{[j]}\left(m_0^2+\frac{j_1^2+\cdots+j_\delta^2}{R^2}\right)
\phi_{[j]}^2\Bigg\}~.
\end{eqnarray}
Here a new scale $m_0$ is the mass of zeroth mode of $\phi$, and modifies the
mass spectrum of $\phi_n$ ($n>0$).
Consequently the spectral function and the two point function changes as
\begin{equation}
\rho_\calU(P^2)=2\pi\sum_n\delta(P^2-m_0^2-m_n^2)F_n^2~,
\label{rhom}
\end{equation}
and
\begin{equation}
D_\calU(P)=\sum_n\frac{iF_n^2}{P^2-m_0^2-m_n^2+i\epsilon}~.
\label{Dm}
\end{equation}
Note that Eqs. (\ref{rhom}) and (\ref{Dm}) are consistent in that
\begin{eqnarray}
D_\calU(P)&=&\sum_n\frac{iF_n^2}{P^2-m_0^2-m_n^2+i\epsilon}\nn\\
&=&
\sum_n\int\frac{dM^2}{2\pi}\frac{i2\pi\delta(M^2-m_0^2-m_n^2)F_n^2}
{P^2-M^2+i\epsilon}\nn\\
&=&
\int\frac{dM^2}{2\pi}\frac{i\rho_\calU(M^2)}{P^2-M^2+i\epsilon}~.
\end{eqnarray}
We have a simple relation between $\rho_\calU^0$ and $\rho_\calU$:
\begin{equation}
\rho_\calU(P^2)=\rho_\calU^0(P^2-m_0^2)~.
\end{equation}
This relation justifies the usual ansatz for the modified spectral function with
the scale-invariance breaking in the literature.
It should be also noted that the term $\sim m_0^2\phi_n^2$ serves as the
infrared cutoff (or a mass gap).
Were it not for the cutoff, the vacuum expectation value of $\calO_\calU$ is
infrared divergent \cite{DEQ}.
The mass gap term was at first put by hand to regulate the IR divergence.
In the language of higher dimensional deconstruction, it is nothing but
the mass of the zeroth mode.
However, the origin of the mass generation is still beyond this framework.
\par
In the $R\to\infty$ limit $\rho_\calU$ is matched onto the continuum spectrum 
much like $\rho_\calU^0$ as in (\ref{rho0match}):
\begin{equation}
\rho_\calU(P^2)\rightarrow A_{d_\calU}\theta(P^0)\theta(P^2-m_0^2)
(P^2-m_0^2)^{d_\calU-2}~.
\label{match}
\end{equation}
From Eqs.\ (\ref{rhom}) and (\ref{match}), one has
\begin{equation}
2\pi\frac{1}{|(m_n^2)'|}F_n^2=A_{d_\calU}(m_n^2)^{d_\calU-2}~,
\end{equation}
where the prime of $m_n^2$ means the derivative with respect to $n$.
Assuming that $(m_n^2)'>0$, one arrives at
\begin{eqnarray}
2\pi F_n^2 dn &=& A_{d_\calU}(m_n^2)^{d_\calU-2}d(m_n^2)~,~~~{\rm or}\\
m_n^2&=&\left[\frac{2\pi}{A_{d_\calU}}(d_\calU-1)\int F_n^2 dn\right]^
\frac{1}{d_\calU-1}~.
\end{eqnarray}
Note that this relation is quite general.
As an example if $F_n\sim n^\alpha$, then 
\begin{equation}
m_n^2\sim n^{(\alpha+1)/(d_\calU-1)}~,
\end{equation}
or conversely, if $m_n^2\sim n^\beta$, then
\begin{equation}
F_n^2\sim n^{\beta(d_\calU-1)-1}~.
\end{equation}
\par
As a simple application, consider the unparticle production process
$t\to u+\calU$.
This was originally considered by \cite{Georgi} and reviewed through
the deconstruction in \cite{Stephanov}, and soon after re-examined with
the IR cutoff in \cite{Fox}.
Here it will be shown that the same result of \cite{Fox} is reproduced by
using the massive $\phi$ in higher dimensions.
\par
Consider first the process of $t\to u+\phi_n$ where $\phi_n$ is treated as
a particle state with mass $\sqrt{m_0^2+m_n^2}$.
The process is governed by the interaction
\begin{equation}
i\frac{\lambda}{\Lambda_\calU^\dU}\ubar\gamma_\mu(1-\gamma_5)t
\partial^\mu\calO_\calU+{\rm H.c.}~,
\end{equation}
whose deconstructed version is
\begin{equation}
i\frac{\lambda}{\Lambda_\calU^\dU}\ubar\gamma_\mu(1-\gamma_5)t
\sum_nF_n\partial^\mu\phi_n+{\rm H.c.}~,
\end{equation}
where $\lambda$ is a dimensionless coupling.
The decay rate $\Gamma_n(t\to u+\phi_n)$ is simply given by a usual two-body
decay process as
\begin{eqnarray}
\Gamma_n(t\to u+\phi_n)&=&
\frac{1}{2m_t}
\int\frac{d^3p_u}{(2\pi)^3}\frac{1}{2E_u}
\int\frac{d^3p_n}{(2\pi)^3}\frac{1}{2E_n}
\left(\frac{1}{2}\sum_{\rm pol.}|\calM|^2\right)
~(2\pi)^4\delta^4(p_t-p_u-p_n)\nn\\
&=&
\frac{|\lambda|^2}{\Lambda_\calU^{2\dU}}\frac{F_n^2}{2\pi}m_tE_u^2~,
\end{eqnarray}
where $\calM$ is the invariant matrix element of the process and 
$p_i$ $(i=t,u,n)$ are the momenta of $t$, $u$, and $\phi_n$ with energies $E_i$.
The $u$ quark energy is definitely given by
\begin{equation}
E_u=\frac{m_t^2-m_n^2-m_0^2}{2m_t}~,
\end{equation}
where $m_t$ is the $t$ quark mass.
For an energy interval $dm_n^2$ there are $dm_n^2/(m_n^2)'$ number of states.
In terms of $dE_u$, one has
\begin{equation}
d\Gamma(t\to u+\calU)=\left[\frac{2m_t}{(m_n^2)'}dE_u\right]
\Gamma_n(t\to u+\phi_n)~.
\end{equation}
From the matching condition of (\ref{match}), 
\begin{equation}
\frac{d\Gamma}{dE_u}
=\frac{|\lambda|^2}{\Lambda_\calU^{2\dU}}\frac{A_\dU}{2\pi^2}m_t^2E_u^2
\big(m_t^2-2m_tE_u-m_0^2\big)^{\dU-2}~,
\label{dGdE}
\end{equation}
which is the same result of \cite{Fox}, since the total decay rate from
(\ref{dGdE}) is 
\begin{equation}
\Gamma=\frac{|\lambda|^2}{\Lambda_\calU^{2\dU}}\frac{A_\dU}{2\pi^2}
\frac{m_t^5(m_t^2)^{\dU-2}}{4\dU(\dU^2-1)}
\left(1-\frac{m_0^2}{m_t^2}\right)^{\dU+1}~.
\end{equation}
Note that (\ref{dGdE}) is also same in form as that of \cite{Stephanov} 
where $m_0=0$.
\par
In summary, we provided a higher dimensional field theory to realize the 
unparticle through deconstruction.
In this framework the higher dimensional mass term $m_0$
breaks the scale invariance of the unparticle sector and plays the role of 
IR cutoff which regulates the IR divergence of the VEV for $\calO_\calU$.
The corresponding spectral function is shifted by an amount of $m_0^2$,
justifying the usual ansatz.
When matching the deconstruction with the continuous spectrum, one gets
a general relation between the decay constant of the deconstructing field
and its mass.
It was also shown that this framework successfully reproduces the decay rate
distribution of unparticle production process.
\par
The origin of $m_0$ is not specified in this work.
It might probably come from the Higgs couplings.
In this case the scalar sector of the theory can contribute to the electroweak
symmetry breaking and change the Higgs mass through mixing \cite{DEQ}.
Whether Higgs is an unparticle (Unhiggs) \cite{Unhiggs} or not is an open 
question.
It will be a good challenge to try deconstruction for the Unhiggs in a
higher dimensional context \cite{Victoria}.


\begin{thebibliography}{99}
\bibitem{Georgi}
 H.~Georgi,
  Phys.\ Rev.\ Lett.\  {\bf 98}, 221601 (2007)
  [arXiv:hep-ph/0703260];
Phys.\ Lett.\  B {\bf 650}, 275 (2007)
  [arXiv:0704.2457 [hep-ph]].
\bibitem{Cheung}
 K.~Cheung, W.~Y.~Keung and T.~C.~Yuan,
  Phys.\ Rev.\ Lett.\  {\bf 99}, 051803 (2007)
  [arXiv:0704.2588 [hep-ph]],
  Phys.\ Rev.\  D {\bf 76}, 055003 (2007)
  [arXiv:0706.3155 [hep-ph]];
M.~Luo and G.~Zhu,
Phys.\ Lett.\  B {\bf 659}, 341 (2008)
  [arXiv:0704.3532 [hep-ph]];
Y.~Liao,
  Phys.\ Rev.\  D {\bf 76}, 056006 (2007)
  [arXiv:0705.0837 [hep-ph]];
T.~Kikuchi and N.~Okada,
  arXiv:0707.0893 [hep-ph];
A.~Lenz,
  Phys.\ Rev.\  D {\bf 76}, 065006 (2007)
  [arXiv:0707.1535 [hep-ph]];
J.~R.~Mureika,
  Phys.\ Lett.\  B {\bf 660}, 561 (2008)
  [arXiv:0712.1786 [hep-ph]];
B.~Grinstein, K.~Intriligator and I.~Z.~Rothstein,
  arXiv:0801.1140 [hep-ph];
N.~V.~Krasnikov,
  Int.\ J.\ Mod.\ Phys.\  A {\bf 22}, 5117 (2007)
  [arXiv:0707.1419 [hep-ph]],
  Mod.\ Phys.\ Lett.\  A {\bf 23}, 3233 (2008)
  [arXiv:0802.0830 [hep-ph]];
F.~Sannino and R.~Zwicky,
  arXiv:0810.2686 [hep-ph];
D.~C.~Dai and D.~Stojkovic,
  arXiv:0812.3396 [gr-qc];
J.-P.~Lee,
  arXiv:0710.2797 [hep-ph],
  arXiv:0803.0833 [hep-ph],
  arXiv:0809.3311 [hep-ph].
\bibitem{BZ}
 T.~Banks and A.~Zaks,
  Nucl.\ Phys.\  B {\bf 196}, 189 (1982).
\bibitem{Fox}
P.~J.~Fox, A.~Rajaraman and Y.~Shirman,
  Phys.\ Rev.\  D {\bf 76}, 075004 (2007)
  [arXiv:0705.3092 [hep-ph]].
\bibitem{Stephanov}
  M.~A.~Stephanov,
  Phys.\ Rev.\  D {\bf 76}, 035008 (2007)
  [arXiv:0705.3049 [hep-ph]].
\bibitem{Terning}
G.~Cacciapaglia, G.~Marandella and J.~Terning,
  arXiv:0804.0424 [hep-ph].
\bibitem{DEQ}
A.~Delgado, J.~R.~Espinosa and M.~Quiros,
  JHEP {\bf 0710}, 094 (2007)
  [arXiv:0707.4309 [hep-ph]].
\bibitem{ACD}
T.~Appelquist, H.~C.~Cheng and B.~A.~Dobrescu,
  Phys.\ Rev.\  D {\bf 64}, 035002 (2001)
  [arXiv:hep-ph/0012100].
\bibitem{Unhiggs}
 D.~Stancato and J.~Terning,
  arXiv:0807.3961 [hep-ph].
\bibitem{Victoria}
 A.~Falkowski and M.~Perez-Victoria,
  arXiv:0810.4940 [hep-ph].
\end{thebibliography}
\end{document}